\documentstyle[12pt,epsfig]{article}
\begin{document}
\baselineskip 0.7cm
%
%
\newcommand{\gsim}{ \mathop{}_{\textstyle \sim}^{\textstyle >} }
\newcommand{\lsim}{ \mathop{}_{\textstyle \sim}^{\textstyle <} }
\def\ul#1{\underline{#1}}
\def\J#1#2{J_{#1}\!\left(#2\right)}
%
\def\rc{r_c}
\def\pikrc{\pi{}k\rc}
\def\z1{e^{\pikrc}}
\def\zc{z_c}
\def\planck5d{M_{{\rm 5d}}}
\def\gauge5d{g_{{\rm 5d}}}
\def\yukawa5d{y_{{\rm 5d}}}
\def\quartic5d{\lambda_{{\rm 5d}}}
%
\def\L{{\rm L}}
\def\R{{\rm R}}
\def\psiL#1{\psi_\L^{(#1)}}
\def\psiR#1{\psi_\R^{(#1)}}
\def\psileft#1#2{\psiL{#1}\!\left(#2\right)}
\def\psiright#1#2{\psiR{#1}\!\left(#2\right)}
%
\def\leftmode#1#2{\xi_{#1}\!\left(#2\right)}
\def\rightmode#1#2{\eta_{#1}\!\left(#2\right)}
\def\gaugemode#1#2{{\widehat \chi}_{#1}\!\left(#2\right)}
\def\zeromode#1{\widehat{\zeta}\!\left(#1\right)}
\def\whleftmode#1#2{{\widehat \xi}_{#1}\!\left(#2\right)}
\def\whrightmode#1#2{{\widehat \eta}_{#1}\!\left(#2\right)}
%
\def\gaugemass#1{M_{{#1}}}
\def\eigen#1{\lambda_{#1}}
\begin{titlepage}
\begin{flushright}
TU-581\\
KEK-TH-665\\
NIIG-DP-99-03\\
December 1999
\end{flushright}

\begin{center}
{\large \bf Bulk Standard Model in the Randall-Sundrum Background}

\vskip 1.2cm

Sanghyeon~{\sc Chang}%
        $^{(a)}$\rlap,\footnote{schang@tuhep.phys.tohoku.ac.jp}
Junji~{\sc Hisano}%
        $^{(b)}$\rlap,\footnote{junji.hisano@kek.jp}
Hiroaki~{\sc Nakano}$^{(c)}%
        $\rlap,\footnote{nakano@muse.hep.sc.niigata-u.ac.jp}
Nobuchika~{\sc Okada}%
        $^{(b)}$\rlap,\footnote{okadan@ccthmail.kek.jp}
 and 
Masahiro~{\sc Yamaguchi}%
        $^{(a)}$\footnote{yama@tuhep.phys.tohoku.ac.jp}

\vspace{10mm}

(a) {\it Department of Physics, Tohoku University,
         Sendai 980-8578, Japan}
\\
(b) {\it Theory Group, KEK, Tsukuba, Ibaraki 305-0801, Japan}
\\
(c) {\it Department of Physics, Niigata University, 
         Niigata 950-2181, Japan}
\vskip 1.5cm

\abstract {
We discuss issues in an attempt to put the Standard Model (SM)
in five-dimensional anti-de Sitter spacetime 
compactified on $S^1/Z_2$.
The recently-proposed approach to the gauge hierarchy problem 
by using this background geometry, with the SM confined on a boundary,
is extended to a situation where (some of)  the SM particles reside
in the five dimensional bulk. 
In particular, we find a localization of zero modes of bulk fermions
near the boundary with a negative tension.
Unlike the compactification with the flat metric, 
these fermion zero modes couple to Kaluza-Klein (KK) excitations 
of the SM gauge bosons.
Interestingly, only low-lying modes of such KK gauge bosons
have non-negligible couplings.
Current electroweak precision data give a constraint that 
the first KK mode be heavier than $9$~TeV.
We also argue that at least the Higgs field should be confined 
on the brane to utilize the Randall-Sundrum background 
as a solution to the gauge hierarchy.
}
\end{center}

\setcounter{footnote}{0}
\end{titlepage}

\section{Introduction}

There have recently been new proposals to the gauge hierarchy problem 
by using  geometry of  extra dimension(s).  
The first of such proposals in Ref.~\cite{ADD} was that 
extra dimensions with large radii can account for 
the weakness of the gravitational interactions in four dimensions, 
even if the fundamental scale is close to the electroweak scale. 
[See also Refs.~\cite{R-S, earlier} for earlier attempts.]

More recently Randall and Sundrum \cite{RS1,GO} 
proposed another approach to the gauge hierarchy 
by utilizing a warped extra dimension. 
In this approach, the spacetime is five dimensional, 
with one extra dimension compactified on $S^1/Z_2$.  
The metric in the Randall-Sundrum (RS) model is 
\begin{equation}
ds^2 = e^{-2 \sigma(y)} \eta_{\mu\nu} dx^{\mu} dx^{\nu} - dy^2 \ ,
\label{BGmetric}
\end{equation}
where $y=x^5$ is a coordinate of $S^1$ with period $2\pi\rc$, 
and $\sigma(y)=k |y|$ with a curvature scale $k$ determined 
by the negative cosmological constant $\Lambda<0$
in the five dimensional bulk.
At each boundary $y=y_i$ ($y_0=0$ and $y_1=\pi\rc$), there locates 
a set of branes, whose tension (vacuum energy) $V_i$ has to be 
fine-tuned to realize four-dimensional Poincar\'e invariance;
\begin{equation}
k^2\equiv\frac{-\Lambda}{24\planck5d^3} \ , \qquad
\frac{V_0}{24\planck5d^3}\,=\,k\,=\,\frac{-V_1}{24\planck5d^3} \ .
\end{equation}
It was then argued that
the Planck mass $M_{{\rm pl}}$ 
in the effective four-dimensional theory is related to
the `fundamental' scale $\planck5d$ in five dimensions by
\begin{equation}
M_{{\rm pl}}^2 = \frac{\planck5d^3}{k}
                 \left(1-e^{-2\pi{}k\rc}\right) \ .
\end{equation}
In the following we assume that 
the both $\planck5d$ and $k$ are of the order $M_{{\rm pl}}$
(with $k\lsim\planck5d$).

The warp factor $e^{- \sigma(y)}$ represents 
 an energy scale of physics phenomena at the position $y$
as measured by the four-dimensional flat metric.
Thus the electroweak scale is naturally realized 
on the distant brane at $y=\pi\rc$ with $V_1<0$ 
if one appropriately adjusts the length of the extra dimension 
to get $k\,e^{-\pikrc} \sim{}100$~--~$1000$ GeV. 
In fact, in the proposal of Ref.~\cite{RS1},
all the Standard Model (SM) particles are {\it assumed}
to be confined on this brane.

Various aspects of this model and 
its extensions \cite{RS2,LR} have been studied 
in the literature \cite{Verlinde,papers1,DHR1,ChangYama,HisanoOkada}.
Among other things, 
Goldberger and Wise pointed out in Ref.~\cite{GW} that 
the physics scale of a scalar field is characterized 
by the warp factor at the distant brane, 
even if it resides in the whole bulk. This leads one to imagine
that the Higgs field can naturally be embedded 
in the bulk of the five dimensional spacetime. 
{}Furthermore the authors of Refs.~\cite{DHR,Pomarol} considered 
the gauge bosons 
in the bulk while the leptons and quarks are on the brane.

In this paper, we would like to pursue this line further,
and in particular consider a situation that 
fermions as well as the gauge bosons  reside in the bulk. 
We will show in section two that 
zero modes of the bulk fermions,
which we identify as quarks and leptons in the SM,
are localized near the brane at $y=\pi\rc$.
This explains why the RS solution to the gauge hierarchy problem
applies also for the bulk SM
even if we are not assuming from the start that
the SM fields are confined on `our' brane;
put differently, the gravity is {\it automatically} weak 
for the matter fields in the bulk SM.
It turns out, however, that such fermions zero modes
couple to Kaluza-Klein (KK) modes of the SM gauge bosons.  
Thus the theory is severely constrained 
by the electroweak measurement because the exchange of 
the KK modes generates four-Fermi type interactions
as we will describe in section four.
This is in contrast to the case with 
the flat metric for the extra dimension, 
where the KK modes of the gauge bosons decouple 
from the zero mode fermions at the tree level \cite{NathYama}.

{}Finally in section five, we will discuss the Higgs mechanism 
and how the gauge bosons and the fermions acquire masses. 
We will mainly examine the simplest case in which
the Higgs field lives also in the bulk and 
develops a {\it constant} vacuum expectation value (VEV).
Then, as is shown in Appendix, the gauge boson masses 
naturally become of the order of the energy scale of our brane, 
which is forced to be much higher than the weak scale 
by the constraint from the current precision experiments, 
unless we make an extreme fine tuning for the Higgs boson mass. 
In this case the gauge hierarchy problem would be back,
and thus the bulk Higgs case should be virtually excluded. 
This leaves the case where the Higgs is confined on our brane.

\section{Bulk Fermion and Localization of Zero Mode}

The five-dimensional Lagrangian
for a free massless fermion $\Psi(x,y)$ can be written as
\begin{eqnarray}
e^{-1}{\cal L}_{{\rm fermion}}
&=& \overline{\Psi}\,i\Gamma^{\ul{A}}\,e_{\ul{A}}{}^A
    \left(\partial_A+\frac{1}{8}\,\omega_A{}^{\ul{B}\,\ul{C}}
                     \left[\Gamma_{\ul{B}},\Gamma_{\ul{C}}\right]
    \right)\Psi \ ,
\end{eqnarray}
where $e_{\ul{A}}{}^A$ is the inverse of the f\"{u}nfbein,
and the gamma matrices in five-dimensions are given by
$\Gamma_{\ul{M}}=\left(\gamma_\mu,\,i\gamma_5\right)$, satisfying 
$\{\Gamma_{\ul{M}},\,\Gamma_{\ul{N}}\}
=2\eta_{\ul{M}\,\ul{N}}=2{\rm diag}\left(+,-,-,-,-\right)$.
In the RS background (\ref{BGmetric}), 
which respects the four-dimensional Poincar\'e invariance,
only non-vanishing component of the spin connection 
$\omega_A{}^{\ul{B}\,\ul{C}}$ is given by
\begin{equation}
\omega_\mu{}^{\ul{\nu}\,\ul{5}}
\,={}-e_\mu{}^{\ul{\nu}}\,e^{\ul{5}\,5}\partial_5\sigma
\,={}+e^{-\sigma}\sigma'\,\delta_\mu{}^{\ul{\nu}} \ ,
\label{connection}
\end{equation}
where $\sigma'=\partial_5\sigma$.
Therefore we obtain
\begin{eqnarray}
{\cal L}_{{\rm fermion}}
 &=& e^{-3\sigma}\overline{\Psi} 
     \left[i\gamma^{\mu} \partial_\mu 
      -\gamma_5\,e^{-\sigma}\left(\partial_5-2\sigma'\right)
     \right]\Psi
\label{L_f}\\
 &=& e^{-\frac{3}{2}\sigma}\overline{\Psi} 
     \left[i\gamma^{\mu} \partial_\mu 
      -\gamma_5\,e^{-\sigma}\left(\partial_5-\frac{1}{2}\sigma'\right)
     \right]e^{-\frac{3}{2}\sigma}\Psi \ .
\nonumber
\end{eqnarray}
Interestingly, the mass operator 
$\gamma_5\,e^{-\sigma}\left(\partial_5-2\sigma'\right)$ for $\Psi$
receives such a piece from the spin connection 
that has {\it a kink profile with a gap}
\begin{equation}
\Delta\sigma'_i
\,\equiv\,\sigma'(y_i+0)-\sigma'(y_i-0)
\,=\,\frac{2V_i}{24\planck5d^3} \ ,
\end{equation}
where $V_i$ is a tension of the brane located at $y=y_i$.
To pursue an analogy with domain wall fermion \cite{Kaplan}
is another motivation to consider the bulk fermions
in the RS background.

Before going into any details,
let us first consider the fermion zero mode 
$\Psi(x,y)=\Psi_0\!\left(x\right)e^{3\sigma(y)/2}\zeromode{y}$
with $i\gamma^\mu\partial_\mu\Psi_0(x)=0$, 
where a factor $e^{3\sigma(y)/2}$ brings 
the kinetic term in Eq.~(\ref{L_f}) into the canonical form.
By solving the five-dimensional Dirac equation,
we find that the zero mode is localized
near the brane with a {\it negative} tension $V_1<0$\,;
\begin{equation}
\zeromode{y}
 = \zeromode{\pi\rc}
   e^{-\frac{k}{2}\left\vert\pi\rc-y\right\vert} \ .
\end{equation}
We should remark that
our mechanism for localizing fermion zero modes quite resembles 
many earlier attempts \cite{defect, R-S, Kaplan, AH-S}
which utilizes a kink background 
induced by a topological defect or scalar field,
except that {\it it is automatic};
our kink mass term in Eq.~(\ref{L_f}) appears not by hand,
but as a consequence of the gravitational background
{\it \`a la}\ Randall and Sundrum.
One may regard the RS background as generated 
by the scalar potential in gauged supergravity \cite{gaugedSUGRA},
but the point we stress here is that
one and the same mechanism is responsible
for the generation of the gauge hierarchy
and the localization of fermions.
%

In the simplest setting we are describing,
the chiral nature of fermions results from the compactification 
on $S^1/Z_2$ by imposing the $Z_2$ projection\rlap.\footnote{
The chiral asymmetry could be produced
if we introduced a suitable Dirac mass term for our bulk fermion.
In fact the five-dimensional parity invariance forbids us 
from introducing such a bare mass term, 
and both chirality of zero modes are localized near the same brane.
}
For the bulk fermion, we impose that 
$\Psi(x,y)$ is even under five-dimensional parity;
\begin{equation}
\gamma_5 \Psi (x,-y)\,={}+\Psi (x,y) \ .
\label{parity}
\end{equation} 
Then there remains only one zero mode with positive chirality 
(right-handed fermion), as we will see shortly.
If we consider the opposite condition
$\gamma_5 \Psi (x,-y)=-\Psi (x,y)$, 
we will have a left-handed fermion as the zero mode.

We make a mode expansion with respect to the fifth dimension;
\begin{equation}
\Psi(x,y) =  
\sum_n \left[\psileft{n}{x} \leftmode{n}{y}
           + \psiright{n}{x}\rightmode{n}{y}\right] \ ,
\end{equation}
where 
$\gamma_5\,\psi_{{\rm L/R}}^{(n)} ={}\mp \psi_{{\rm L/R}}^{(n)}$.
Using this expansion in Eq.~(\ref{L_f}) and integrating over $y$,
we get the four-dimensional effective theory
\begin{equation}
{\cal L}^{({\rm 4dim})}_{{\rm fermion}}
 = \sum_n \left[\,
   \overline{\psiL{n}} i \gamma^\mu\partial_\mu \psiL{n}
  +\overline{\psiR{n}} i \gamma^\mu\partial_\mu \psiR{n}
  -\left(m_n \overline{\psiL{n}}\psiR{n}+ {\rm H.c.}\right) 
   \right] \ .
\end{equation}
Here the mode functions satisfy the eigenvalue equations
\begin{eqnarray}
{}-e^{-\sigma}\left(\partial_y-2\sigma'\right) \leftmode{n}{y}
 &=& m_n\,\rightmode{n}{y} \ ,
\label{EOM:left}
\\
{}+e^{-\sigma}\left(\partial_y-2\sigma'\right)\rightmode{n}{y}
 &=& m_n\, \leftmode{n}{y}
\label{EOM:right}
\end{eqnarray}
with the normalizations
\begin{equation}
\int^{\pi\rc}_0 dy\,e^{-3 \sigma}
  \leftmode{n}{y} \leftmode{m}{y}
= \int^{\pi\rc}_0 dy\,e^{-3 \sigma}
  \rightmode{n}{y}\rightmode{m}{y}
= \delta_{mn} \ .
\label{normalization:fermion}
\end{equation}
Since the condition (\ref{parity}) is translated into
$\leftmode{n}{y}=\!{}-\leftmode{n}{-y}$ and 
$\rightmode{n}{y}=\!{}+\rightmode{n}{-y}$,
the $Z_2$ projection and the periodicity
$\Psi(x, y+2\pi\rc)=\Psi(x, y)$ give the boundary conditions
\begin{eqnarray}
\leftmode{n}{y\!=\!y_i}\,=\,0\,=\,\partial_y\rightmode{n}{y\!=\!y_i}
\end{eqnarray}
at $y_0=0$ and $y_1=\pi\rc$.
With these conditions,
one can easily find the explicit solution for the mode functions.
We present the result for 
$\whleftmode{n}{y}\equiv{}e^{-3\sigma(y)/2}\leftmode{n}{y}$
and $\whrightmode{n}{y}\equiv{}e^{-3\sigma(y)/2}\rightmode{n}{y}$,
for which a physical picture is most transparent
(since the normalization conditions (\ref{normalization:fermion}) 
becomes the canonical ones\footnote{
This is similar to the rescaling 
that was discussed in Ref.~\cite{DDG},
but it is not exactly the same because we are considering
the effective theory after integrating over the fifth-dimension,
not that on the brane.
}\,)\,;
\begin{eqnarray}
\whleftmode{n}{y}
 &=& \sqrt{\frac{2 k}{1-e^{-\pi{}k\rc}}}\,
     e^{-\frac{k}{2}|\pi\rc-y|}
     \sin\frac{m_n}{k}\left(e^{\sigma(y)}-1\right) \ ,
\nonumber\\
\whrightmode{n}{y}
 &=& \sqrt{\frac{2 k}{1-e^{-\pi{}k\rc}}}\,
     e^{-\frac{k}{2}|\pi\rc-y|}
     \cos\frac{m_n}{k}\left(e^{\sigma(y)}-1\right)
\end{eqnarray}
for $m_n = n \pi k/(e^{\pikrc}-1)\neq{}0$. 
{}For the zero mode $m_0=0$, 
\begin{equation}
\whleftmode{0}{y}  = 0, \qquad
\whrightmode{0}{y} = \sqrt{\frac{k}{1-e^{-\pi{}k\rc}}}\,
                     e^{-\frac{k}{2}|\pi\rc-y|} \ .
\label{zeromode}
\end{equation}
This clearly shows that 
the right-handed fermion zero mode is localized
near the orientifold plane at $y=\pi\rc$ 
while the left-handed zero mode is projected out.

As mentioned above,
the left-handed zero mode can be obtained by the opposite projection.
One expects that, as in the SM, these fermion zero modes will 
acquire their masses 
through Yukawa couplings to the Higgs field.
To realize this in our model, we prepare 
an $SU(2)$ doublet $\Psi_\L(x,y)$ and a singlet $\Psi_\R(x,y)$,
and impose the $Z_2$-projection condition 
$\gamma_5\Psi_{{\rm L/R}}(x,-y)=\!{}\mp \Psi_{{\rm L/R}}(x,y)$.
Then the $Z_2$-invariant operators are 
given by $\overline{\Psi}_\R\Psi_\L{}H$.
As for the Higgs field $H$,
there are two distinct possibilities that 
$H$ lives also in the bulk, or it is confined on the brane at $y=\pi\rc$.
Which of two cases leads to a viable model 
is the subject of the subsequent sections.

\section{Gauge Bosons in  the Bulk}
Let us now proceed to the bulk gauge bosons,
which were recently discussed in Refs.~\cite{DHR,Pomarol}. 
Here we briefly discuss the abelian case for simplicity.
The Lagrangian for a bulk gauge field 
in the RS background (\ref{BGmetric}) is given by 
\begin{eqnarray}
{\cal L}_{{\rm gauge}}
 ={}-\frac{1}{4}\left(F_{\mu\nu}\right)^2
    +e^{-2\sigma}\left[\,
      \frac{1}{2}\left(\partial_5{}A_\mu\right)^2
     -\partial_5{}A_\mu\partial^\mu{}A_5
     +\frac{1}{2}\left(\partial_\mu{}A_5\right)^2\,\right] \ ,
\end{eqnarray}
where the contraction by using the flat metric should be understood.
The action principle requires a gauge-invariant boundary condition
$0=F_{5\mu}=\partial_5{}A_\mu-\partial_\mu{}A_5$
at $y_0=0$ and $y_1=\pi\rc$,
but $Z_2$-orbifold projection implies stronger conditions
\begin{eqnarray}
\partial_5{}A_\mu\!\left(x,y\!=\!y_i\right)
\,=\,0\,=\,A_5\!\left(x,y\!=\!y_i\right) \ .
\end{eqnarray}
That is, 
$Z_2$ projection implies the Neumann (Dirichlet)-type
boundary condition for $A_\mu$ ($A_5$).
Although we can proceed in a gauge covariant manner\rlap,\footnote{
In this case,
the `Nambu-Goldstone' field $A_5(x,y)$ should be mode expanded
by using the Bessel functions of the order $\nu=0$
to diagonalize the mixing between $A_\mu^{(m)}(x)$ and $A_5^{(n)}(x)$.
Note also that the zero mode of $A_5$ is projected out.
}
let us take $A_5=0$ gauge \cite{DHR} for simplicity.
Then after integrating by parts, the Lagrangian reduces to
\begin{equation}
{\cal L}_{{\rm gauge}}
 = {}-\frac14\left(F_{\mu\nu}\right)^2
      -\frac1{2}\,A^\mu
       \partial_5\!\left(e^{-2\sigma}\partial_5 A_\mu\right) \ ,
\label{L_g}
\end{equation}
supplemented with Gauss law constraint 
$0\approx\partial_5\left(\partial^\mu{}A_\mu\right)$.

Let us expand $A_\mu$ into the KK modes as
\begin{equation}
A_\mu(x,y) =  
\sum_n A_\mu^{(n)}\!\left(x\right)\gaugemode{n}{y} \ ,
\end{equation}
by using mode functions $\gaugemode{n}{y}$ specified by the conditions
\begin{eqnarray}
{}-\partial_y \left(e^{-2\sigma} \partial_y \gaugemode{n}{y}\right) 
&=& \gaugemass{n}^2\,\gaugemode{n}{y} \ ,
\\
\int^{\pi\rc}_0 dy\, \gaugemode{n}{y}\gaugemode{m}{y}
&=& \delta_{mn} \ ,
\end{eqnarray}
as well as Neumann-type boundary condition
$\partial_y \gaugemode{n}{y_i} = 0$ at $y_0=0$ and $y_1=\pi\rc$.
Substituting this expansion into Eq.~(\ref{L_g})
and integrating over $y$ gives the four-dimensional effective theory
\begin{equation}
{\cal L}_{{\rm gauge}}^{({\rm 4dim})}
 = \sum_n\left[
   {}-\frac{1}{4}\left(F_{\mu\nu}^{(n)}\right)^2
     +\frac{1}{2}\,\gaugemass{n}^2 A_\mu^{(n)}A^{(n)\mu}\right] \ .
\end{equation}
The explicit form of $\gaugemode{n}{y}$ is given by
the Bessel functions of the order $\nu=1$;
\begin{equation}
\gaugemode{n}{y}
= \frac{\sqrt{2k}\,e^{\sigma(y)}}{N_n} 
  \left[     J_1\!\left(\eigen{n}e^{\sigma(y)}\right)
        +c_n Y_1\!\left(\eigen{n}e^{\sigma(y)}\right)
  \right] \ ,
\label{mode:gauge}
\end{equation}
where $\eigen{n}\equiv{}\gaugemass{n}/k\neq{}0$, 
and by denoting $\zc\equiv{}e^{\pikrc}$,
\begin{eqnarray}
c_n
\,={}-\,\frac{J_0\!\left(\eigen{n}\right)}
             {Y_0\!\left(\eigen{n}\right)} \ , \qquad
N_n^2 
&=& \int_1^{\zc}2zdz\Big[J_1\!\left(\eigen{n}z\right) 
                     +c_n Y_1\!\left(\eigen{n}z\right)\Big]^2
\nonumber\\
&=& \left.z^2
    \Big[J_1\!\left(\eigen{n}z\right) 
     +c_n Y_1\!\left(\eigen{n}z\right)\Big]^2
    \right|_1^{\zc} \ .
\label{norm:gauge}
\end{eqnarray}
The mass eigenvalues are determined 
by the condition $\partial_y \gaugemode{n}{\pi\rc}=0$\,;
\begin{equation}
J_0\!\left(\eigen{n}\right)
Y_0\!\left(\eigen{n}e^{\pi{}k\rc}\right)
=
Y_0\!\left(\eigen{n}\right)
J_0\!\left(\eigen{n}e^{\pi{}k\rc}\right) \ .
\end{equation}
The behaviour of the mass eigenvalues $\gaugemass{n}$
is depicted in Fig.~1, where we plot the values of 
$\left(\gaugemass{n}/k\right)\exp\left[\sigma(\pi\rc)\right]$
for $n=1,\cdots,40$.
Asymptotically at higher mass level $n\gg{}1$,
the mode functions behave like
\begin{equation}
\gaugemode{n}{y}
\,\sim{}\,\sqrt{\frac{2k}{1-e^{-\pi{}k\rc}}}\,
           e^{-\frac{k}{2}|\pi\rc-y|}
           \cos\left(n\pi\frac{e^{\sigma(y)}-1}{e^{\pikrc}-1}\right) \ .
\label{gaugemode:as}
\end{equation}
with the same mass eigenvalues $\gaugemass{n}\sim{}m_n$
as the KK fermion masses.
The zero mode is flat in the extra dimension,
$\gaugemode{0}{y}=1/\sqrt{\pi\rc}$,
and the KK gauge bosons show the universal behaviour
of localizing near the brane at $y_1=\pi\rc$
as in other bulk fields.

\section{Bulk Phenomenology}

In this section we will examine phenomenological constraints on the
bulk gauge bosons and fermions. For the moment we assume that a Higgs
mechanism takes place and the zero-modes corresponding to the W and Z
bosons acquire tiny masses of the weak scale. We will discuss the detail
of the mechanism in the next section.

In the case that both fermions and gauge bosons are living in the bulk,
the gauge coupling of the bulk fermion to the bulk gauge boson 
is written as
\begin{equation}
e^{-1}{\cal L}_{{\rm coupling}}
 = \gauge5d\overline{\Psi}\!\left(x,y\right)
   i\Gamma^{\ul{M}}\,e_{\ul{M}}{}^{M}\!\left(y\right)
   A_{M}\!\left(x,y\right)\Psi\!\left(x,y\right) \ .
\end{equation}
Using the results given above, we find that
the coupling constant of a KK mode of the gauge boson 
to the massless (zero-mode) fermion bilinear is given by
\begin{eqnarray}
g_n &=& g\,\frac{\sqrt{2\pi{}k\rc}}{N_n}
        \int_1^{\zc}\,\frac{zdz}{\zc-1}
        \Big[J_1\!\left(\eigen{n}z\right)
        +c_n Y_1\!\left(\eigen{n}z\right)\Big] \ ,
\label{KKcouling}
\end{eqnarray}
where $\zc=e^{\pikrc}$, 
and $g=\gauge5d/\sqrt{\pi\rc}$ is the four-dimensional
gauge coupling constant. In Fig.~2, we plot the values of $g_n/g$.
We found that the KK modes of the gauge boson have
non-vanishing couplings to the bilinear of the zero-mode fermions.  
This is in sharp contrast to the flat metric case 
(or the factorizable extra dimension), where the conservation of 
the fifth-dimensional momentum prohibits these couplings.
Another interesting point to be stressed is that 
{\it only the first KK mode of the gauge boson strongly couples 
to the fermion zero-mode.}  We find
\begin{equation}
 \frac{g_1}{g}\ \simeq\ 4.1  \ , \qquad
 \frac{g_2}{g}\ \simeq\ 0.55 \ , \qquad
 \frac{g_3}{g}\ \simeq\ 0.54 \ ,
\label{eq:gn/g}
\end{equation}
and $g_n \ll 1$ for higher $n$. 
Physically this suppression for higher KK modes is understood 
by the oscillating behaviour (\ref{gaugemode:as})
of the Bessel functions. 
Thus one may expect that 
the high energy behaviour of this model is rather moderate. 
Note that this is quite different from the case of the brane fermion 
where the coupling is determined by the wave function at the brane 
and turns out to be universal,
i.e. $g_n/g=\sqrt{2\pi{}k\rc} \simeq 8.4$ for all KK modes.

Phenomenologically the existence of the non-vanishing couplings 
(\ref{KKcouling}) plays an important role \cite{flat,flat2} 
because the exchange of the KK modes of the gauge bosons 
induces four-Fermi interactions. 
{}For the weak boson case,
following Ref.~\cite{DHR}, it is convenient to define
\begin{equation}
V\,\equiv\,\frac{\sum g_n^2/\gaugemass{n}^2}{g^2/m_W^2}
\,=\,\sum_{n=1}^{\infty}\left(\frac{g_n}{g}\right)^2
                        \frac{m_W^2}{\gaugemass{n}^2} \ .
\end{equation}
Using (\ref{eq:gn/g}), we approximate the above equation to 
\begin{equation}
V_{{\rm bulk}} 
\ \approx\ 4.1^2\,\frac{m_W^2}{\gaugemass{1}^2}
\ \approx\ 17  \,\frac{m_W^2}{\gaugemass{1}^2} \ .
\label{eq:Vbulk}
\end{equation} 
Here it is interesting to compare it 
with the case of the brane fermion.
In this case, as we mentioned, $g_n/g=\sqrt{2\pi{}k\rc} \simeq 8.4$ 
for all $n$, and $\sum \gaugemass{1}^2/\gaugemass{n}^2 \simeq 1.5$, 
we find
\begin{equation}
V_{{\rm brane}}
\ \approx\ 8.4^2 \times 1.5\, \frac{m_W^2}{\gaugemass{1}^2}
\ \approx\ 100\,\frac{m_W^2}{\gaugemass{1}^2} \ .
\label{eq:Vbrane}
\end{equation}
Comparison between Eq.~(\ref{eq:Vbulk}) and Eq.~(\ref{eq:Vbrane}) 
implies that, for a given experimental constraint on $V$, 
the bound on the first excited mode in the bulk fermion case
is weaker than that in the brane fermion case 
by a factor $\sqrt{100/17} \sim  2.5$. 

Using the data of the electroweak precision measurements, 
Ref.~\cite{DHR} gives the constraint $V<0.0013$ at 95\% C.L. 
In our case of the bulk fermion, this gives the following bound 
on the mass of the first KK excitation of the W-boson;
\begin{equation}
  \gaugemass{1}\ \gsim\ 9\ \mbox{TeV} \ .
\end{equation}
Note that this bound is certainly weaker than the case of 
Ref.~\cite{DHR}, though it still exceeds the electroweak scale.

Another stringent bound comes from photon and gluon. 
The KK modes of the photon and gluon will 
effectively generate contact interactions 
\begin{equation}
{\cal L}_{{\rm eff}}\,=\,\frac{2 \pi}{ \Lambda^2} J^\mu J_\mu \ .
\end{equation}
Experimentally $\Lambda$ is constrained 
to be higher than $2$~--~$4$ TeV \cite{PDG}, 
with detail depending on which current one considers.  
Note that the coupling of the first KK mode is enhanced by $g_1=4.1g$.
Thus this constraint alone will raise the bound 
on the first excited state well above $1$~TeV.

In passing, some remarks are in order.
{}Firstly, the reason for having so stringent constraints is that
the first KK mode couples to fermions more strongly 
than the massless gauge boson; recalling Eq.~(\ref{norm:gauge}),
we can approximate Eq.~(\ref{KKcouling}) 
for a large $\zc=e^{\pikrc}$ to
\begin{equation}
\frac{g_1}{g}
\,\approx\,\sqrt{2\pikrc}\int_1^{\zc}\,\frac{zdz}{\zc^2}\,
           \frac{J_1\!\left(\eigen{1}z\right)}
                {J_1\!\left(\eigen{1}\zc\right)}
\,\sim\,\frac{\sqrt{2\pikrc}}{2} \ .
\nonumber
\end{equation}
This fact can be understood by noting that although 
the zero mode of the gauge boson is flat in the fifth dimension,
the first KK mode is localized (without oscillating)
near the TeV brane where fermion zero modes are also localized.
Secondly, we comment on how the constraint could be changed
when we consider the massive gauge bosons.
In that case, as we describe in the next section,
the lowest mode of a massive bulk gauge boson
has the mass of the order $\gaugemass{1}$
(unless we make an extreme fine tuning of the bulk gauge boson mass).
Given that, one might wonder whether
the gauge coupling of our W-boson should be identified
with $g_1$, not $g$ of the zero mode,
and it would be $M_2$, not $M_1$, that
should be constrained as the mass of the first KK mode.
If this were the case, the constraint discussed above
would have further been relaxed by a factor $g_1/g_2\sim{}7.5$\,;
$M_2/m_W\approx{}M_2/M_1\sim\left(g_2/g_1\right)/\sqrt{V}\,\gsim{}3.7$.
%
%
Unfortunately, however,
this is actually ruled out
from another constraint coming from the KK photon and gluons.

\section{Higgs Mechanism and Gauge Boson Mass}

Now, we would like to discuss the mechanisms to generate 
the gauge boson mass. Let us first consider the case that 
the Higgs boson is also in the bulk. If we assume that 
the potential of the five-dimensional Higgs field takes the form
\begin{equation}
V(H)={}-\mu^2 H^{\dagger} H 
       +\frac{\quartic5d}{2}\left(H^{\dagger} H\right)^2
\end{equation}
with a negative mass-squared, the Higgs field develops 
the constant VEV in the bulk\footnote{
This VEV should be sufficiently smaller than the curvature scale $k$
not to disturb the background;
otherwise, it could be an origin of the bulk vacuum energy $\Lambda$.
} 
$\sim\sqrt{\mu^2/\quartic5d}$,
which generates the bulk mass term $m$ for the gauge boson. 
Then the mode functions are expressed like in Eq.~(\ref{mode:gauge}) 
but with the order $\nu=\sqrt{1+m^2/k^2}$. 
With the constraint $k\,e^{-\pikrc}$ of the order $10$~TeV or higher,
one has to take a small mass parameter $m$
to realize the gauge boson mass of $100$~GeV.

One might naively expect that
a moderate fine tuning of $m/k\sim{}10^{-2}$
would be enough to realize the correct gauge boson mass
since there would be an approximate zero mode
for a small bulk mass $m$.
In fact, as we show explicitly in Appendix,
the lowest mass eigenvalue $\gaugemass{1}'$ for a very small $m$
is proportional to $m$\,;
\begin{equation}
\gaugemass{1}^{\prime\,2}
\,\simeq\,\frac{m^2}{2 \ln e^{\sigma(y_1)}}
\,=\,\frac{m^2}{2\pi k r_c} \ ,
\label{mass:lowest}
\end{equation}
but there is no suppression by a warp factor!

The absence of a warp factor and the fate of the zero mode may be 
understood by regarding the small bulk mass $m$ as a perturbation;
evaluating the bulk mass term by using the zero mode eigenfunction 
$\gaugemode{0}{y}=1/\sqrt{\pi{}k}$ in the massless case,
we find
\begin{equation}
\gaugemass{1}^{\prime\,2}
\,\simeq\,m^2\int_0^{\pi\rc}dy\,e^{-2\sigma(y)}
          \gaugemode{0}{y}\gaugemode{0}{y}
\,=\,\frac{m^2}{2\pikrc}\left(1-e^{-2\pikrc}\right) \ .
\end{equation}
This will be a good approximation
to the exact mass eigenvalue $\gaugemass{1}'$ 
as far as the mixings between the `zero mode' 
$A^{(0)}_\mu\!\left(x\right)$ and `non-zero modes' 
$A^{(n)}_\mu\!\left(x\right)$ are small
\begin{equation}
\gaugemass{0n}^2
\,=\,m^2\int_0^{\pi\rc}dy\,e^{-2\sigma(y)}
     \gaugemode{0}{y}\gaugemode{n}{y}
\,\ll\,\gaugemass{1}^2 \ .
\end{equation}
When the bulk mass (and thus the mixings) goes up
and becomes comparable to $\gaugemass{1}$,
then the perturbation breaks down and we will find that
the lowest mass eigenvalue $\gaugemass{1}'$ smoothly goes up 
and eventually becomes of the same order as $\gaugemass{1}$
of the first excited state in the massless case.
Apparently the (approximate) zero mode disappears
even for, say, $m/k\sim{}10^{-10}$.

Therefore the mass parameter $m$ itself
must be much smaller than $k \sim{}M_{{\rm pl}}$ 
whereas the natural value for $m$ would be of the order $k$. 
Since the constraints discussed in the previous section push the 
energy scale $k\,e^{-\pikrc}$ of our brane well above $1$~TeV, 
the mass parameter $m$ must be chosen to be the electroweak scale.
This small $m$ parameter for the gauge boson requires 
a hierarchically small $\mu$ parameter in the Higgs potential. 
This is nothing but the conventional fine tuning of 
the Higgs mass in non-supersymmetric theories
and the gauge hierarchy is not solved at all. 
Thus we should discard the model with the bulk Higgs mechanism.

This leaves the case where 
the Higgs is confined on our brane\rlap.\footnote{
Another logical possibility would be
to consider the bulk Higgs field with a positive mass-squared
and to expect that some dynamics 
(in four dimensional effective theory) 
would drive the mass-squared of its lowest mode negative.
}
In this case, the energy scale of the brane is already reduced 
to be $k\,e^{-\pikrc}\sim 10$ TeV. 
Thus to realize the electroweak scale,
the Higgs mass parameter should be tuned just by $10^{2}$.  
This should be compared with the previous case of 
the bulk Higgs mechanism where we need the conventional $10^{16}$. 
In fact, the brane Higgs seems to be the only choice we can take 
to avoid the extreme fine tuning of the Higgs mass
in the `bulk SM' approach.

\section{Conclusions}

We have discussed various issues in an attempt 
to construct bulk Standard Model in the RS background geometry.
In particular, 
by solving the Dirac equation in this background geometry,
we observed the localization of bulk fermion
due to the kink profile of the spin connection.
Since the localization takes place
near the brane with a negative tension where the gravity is weak,
the bulk SM makes the RS approach to the hierarchy problem
more attractive.
The chiral nature of the fermion zero mode is realized
by the $Z_2$-orbifold projection in the present model.

We have also found that the couplings of fermion zero modes 
to the (oscillating) KK modes of the gauge boson 
are suppressed compared with the brane fermion case.  
This relaxes the phenomenological constraint, but not enough. 
In fact the first KK mode of the $W$ gauge boson must be heavier 
than $9$~TeV, which implies that the energy scale 
of the distant brane itself must exceed the TeV scale.

With this phenomenological constraint, 
the bulk Standard Model suffers from the fine-tuning problem.
In particular, when whole the Standard Model is put in the bulk
as we discussed in the last section,
the hierarchy problem is not solved at all and 
we need an extreme fine tuning to realize the electroweak scale. 
In this case the RS background has nothing to do with 
the hierarchy problem, and we need completely another mechanism, 
for instance supersymmetry, to realize the idea of the bulk SM.

If we want to keep the advantage of the RS setting
as a solution of the hierarchy problem, 
we have to confine the Higgs field on the TeV brane.
In this case, the VEV of the Higgs localized at the brane will give 
contributions to the masses of the gauge bosons and fermions. 
We can easily construct a viable model which 
contains the Standard Model particles as the lowest modes
once we accept a moderate fine tuning of $1/100$.
Of course,
some care should be taken of to ensure the proton stability;
higher-dimensional operators will be suppressed
only by the mass scale of the TeV brane
and should be forbidden by some symmetry reasons for instance.
In this respect,
the bulk SM in the simplest formulation suffers from 
the similar problem as in models with large extra dimensions.

Besides phenomenological implications, 
the present formulation of bulk fermion in the RS background 
(and its generalization) will be deserve further study.  
Among others, an interesting application would be 
to formulate chiral fermions on a lattice.

\vskip 5mm

While we have completed our manuscript, 
there appeared interesting preprints \cite{GN,CDH};
the former deals with right-handed neutrinos in the bulk,
and the latter discusses the dynamical Higgs scenario
in the extra-dimension(s).

\section*{Acknowledgments}

We thank the Summer Institute 99 at Yamanashi, Japan,
where the present work was initiated. 
SC, HN and MY also acknowledge KEK for its kind hospitality
where a part of the work was done. 
HN thanks T.~Kugo and H.~So for discussions.
This work was supported in part 
by the Grant-in-Aid for Scientific Research from the Ministry of 
Education, Science, Sports and Culture of Japan, on Priority Area 707
``Supersymmetry and Unified Theory of Elementary Particles" 
(JH and MY), 
and by the Grant-in-Aid No.10740133 (JH), No.11640246 (MY),
and No.98270 (SC).  SC and NO thank the Japan Society
for the Promotion of Science for financial support.

\section*{Appendix: Fate of the Boson Zero Modes}

Here we discuss how the masses of the lowest modes
for spin 0 and 1 particles behave
when they have a non-zero bulk mass $m$.

As usual, the $n$-th mode of a bosonic bulk field
is expressed in terms of Bessel functions as
\begin{equation}
\chi_n\!\left(y\right)
\,=\,\frac{\sqrt{2k}}{N_n}\left(\frac{x_n}{2}\right)^a
     \left[\J{\nu}{x_n}+\alpha_n \J{-\nu}{x_n}\right] \ ,
\end{equation}
where
$N_n$ and $\alpha_n$ are proper normalization constants
and $x_n =\left(\gaugemass{n}'/k\right)e^{\sigma(y)}$.
The order $\nu$ is given by
\begin{equation}
\nu\,=\,\sqrt{a^2+\frac{m^2}{k^2}}\,\simeq\,a+\Delta \nu \ ,
\end{equation}
where $a=2$ for scalar and $a=1$ for vector boson.
$\Delta \nu=0$ corresponds to the vanishing bulk mass.

We are interested in the mass eigenvalue
$\lambda_1\equiv\gaugemass{1}'/k$ of the lowest mode
$\chi_1\!\left(y\right)$. Let us consider the situation
in which the bulk mass $m$ is small enough that the resulting mass
is tiny $\left(\gaugemass{1}'/k\right)e^{\pikrc}\ll{}1$.
Then we can make the approximation for the Bessel functions
near the origin; for $x_1\ll{}1$,
\begin{eqnarray}
\J{\nu}{x_1}
&\simeq& \left(\frac{x_1}{2}\right)^\nu\frac{1}{\Gamma(1+\nu)} \ ,
\nonumber\\
\J{-\nu}{x_1}
&\simeq& \left(\frac{x_1}{2}\right)^{-\nu}
         \left[\frac{1}{\Gamma\!\left(1-\nu\right)}
              -\frac{1}{\Gamma\!\left(2-\nu\right)}
               \left(\frac{x_1}{2}\right)^2\,
         \right] \ .
\nonumber
\end{eqnarray}
At $y=0$, $x_1=\lambda_1$ and the boundary condition gives
\begin{eqnarray}
{}-\frac{1}{\alpha_1}
 &=& \left.
     \frac{\frac{d}{d x}
           \left[\,\left(x/2\right)^a \J{-\nu}{x}\,\right]}
          {\frac{d}{d x}
           \left[\,\left(x/2\right)^a\, \J{\nu}{x}\ \right]}
     \right\vert_{x=\lambda_1}
\nonumber\\
&\simeq& \left(\frac{\lambda_1}{2}\right)^{-2\nu}
         \frac{\Gamma(1+\nu)}{a+\nu}
         \left[\frac{a-\nu}{\Gamma\!\left(1-\nu\right)}
              -\frac{a+2-\nu}{\Gamma\!\left(2-\nu\right)}
               \left(\frac{\lambda_1}{2}\right)^2
         \right] \ .
\end{eqnarray}
The boundary condition at the other boundary $y=\pi\rc$ gives
\begin{equation}
{}-\frac{1}{\alpha_1}
\,\simeq\,\left(\frac{\lambda_1\zc}{2}\right)^{-2\nu}
          \frac{\Gamma\!\left(1+\nu\right)}{a+\nu}
          \left[\frac{a-\nu}{\Gamma\!\left(1-\nu\right)}
               -\frac{a+2-\nu}{\Gamma\!\left(2-\nu\right)}
                \left(\frac{\lambda_1\zc}{2}\right)^2
          \right] \ ,
\end{equation}
where $\zc = e^{\pikrc}$.
These two equations can be summarized as
\begin{eqnarray}
\frac{a+2-\nu}{\Gamma\!\left(2-\nu\right)}
\left(\frac{\lambda_1 }{2}\right)^2
\,=\,\frac{a-\nu}{\Gamma\!\left(1-\nu\right)}\,
     \frac{1- \zc^{-2\nu}}{1- \zc^{2(1-\nu)}} \ ,
\nonumber
\end{eqnarray}
which leads a relation
\begin{equation}
\lambda_1^2
\,\simeq\,\frac{2\left(\nu-1\right)}{1- z_c^{2(1-\nu)}}\,\Delta\nu
          \ .
\label{relation}
\end{equation}
{}For the scalar case, $\nu\simeq{}a=2$,
this correctly reproduces the result in Ref.~\cite{GW};
\begin{equation}
\lambda_1
\,=\,\frac{\gaugemass{1}'}{k}
\,\simeq\,\sqrt{2 \Delta\nu}
\,\simeq\,\frac{1}{\sqrt{2}}\,\frac{m}{k} \ .
\label{mass:lowest:scalar}
\end{equation}
On the other hand,
for the vector boson case, $\nu\simeq{}a =1$,
Eq.~(\ref{relation}) reduces to
\begin{equation}
\lambda_1^2
\,\simeq\,\frac{2 \Delta\nu\left(\nu-1\right)}
               {1-\left(1+2\left(1-\nu\right)\ln z_c\right)}
\,\simeq\,\frac{\Delta\nu}{\ln z_c} \ ,
\end{equation}
which gives the announced relation (\ref{mass:lowest});
\begin{equation}
\frac{\gaugemass{1}'}{k}
\,\simeq\,\frac{1}{\sqrt{2\pikrc}}\,\frac{m}{k} \ .
\label{mass:lowest:gauge}
\end{equation}
We note again that these results
(\ref{mass:lowest:scalar}) and (\ref{mass:lowest:gauge}),
with no suppression by a warp factor,
are valid only for a sufficiently small bulk mass $m$.

%
%
\newcommand{\Journal}[4]{{\sl #1} {\bf #2} {(#3)} {#4}}
\newcommand{\PL}{Phys.~Lett.}
\newcommand{\PR}{Phys.~Rev.}
\newcommand{\PRL}{Phys.~Rev.~Lett.}
\newcommand{\NP}{Nucl.~Phys.}
\newcommand{\ZP}{Z.~Phys.}
\newcommand{\PTP}{Prog.~Theor.~Phys.}
\newcommand{\NC}{Nuovo~Cimento}
\newcommand{\PRP}{Phys.~Rep.}

\clearpage

%
%
\begin{figure}[p]
\centerline{\leavevmode\psfig{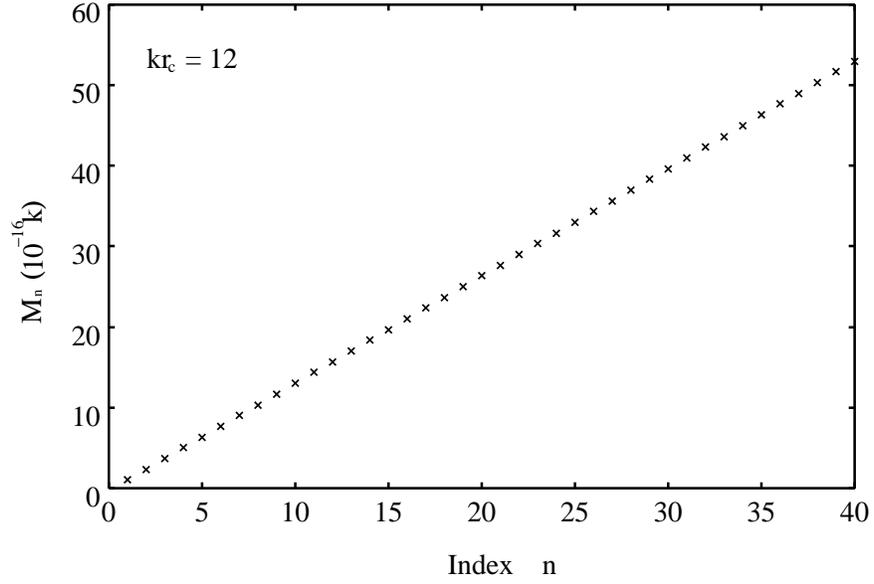}}
\caption{
Plot of the masses of the $n$-th Kaluza-Klein modes of 
the gauge bosons
in units of $ 10^{-16}k$. We take $k\rc=12$. 
}
\end{figure}
%
%
%
%
\begin{figure}[p]
\centerline{\leavevmode\psfig{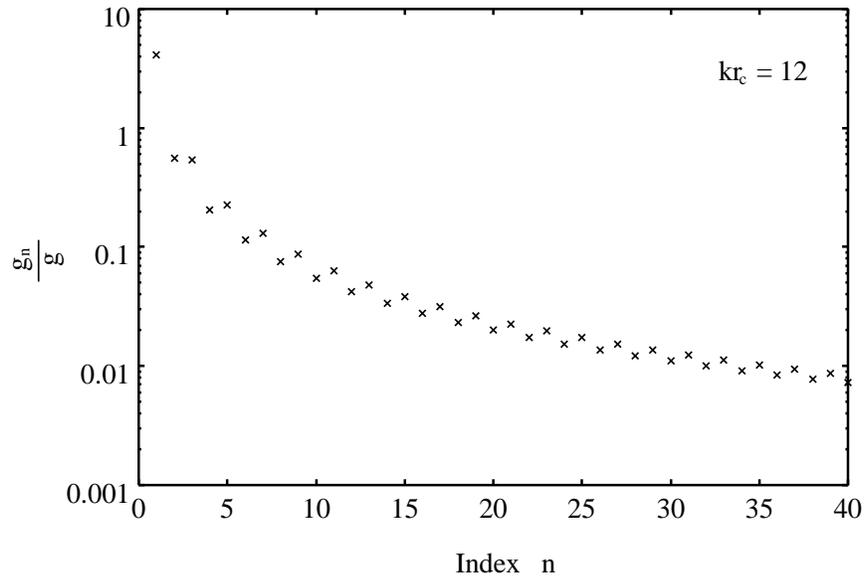}}
\caption
{
Plot of the couplings $g_n$ of the $n$-th Kaluza-Klein modes 
to the bilinear
of the zero modes of the bulk fermions relative to the gauge
coupling constant $g$.
}
\end{figure}
\end{document}